# Structural Analysis of Metal-doped Calcium Oxalate

Eva Weber[1], Davide Levy[1], Matanya Ben Sasson[1], Andy N. Fitch[2] and Boaz Pokroy[1]

[1] Department of Materials Science and Engineering and the Russell Berrie Nanotechnology Institute, Technion−Israel Institute of Technology, 32000 Haifa, Israel
[2] European Synchrotron Radiation Facility, B.P. 220, 38043 Grenoble Cedex, France

**Abstract**

Calcium oxalate crystals are the most common biominerals found in plants. They also make their presence known as painful kidney stones in humans and animals. Their function in plants is extraordinarily versatile and encompasses calcium storage and defense mechanisms against herbivores and detoxification processes. Since plants containing calcium oxalate crystals are often exposed to metallic environments, we studied the interaction of such environmental metals with calcium oxalate *in vitro*. We showed that selected metals are indeed incorporated into synthetic calcium oxalate, and in a manner that depends on their ionic radius when precipitated in solution. One such mechanism of incorporation is based on the replacement of calcium ions by other metal cations within the host unit cell. The unit cell of calcium oxalate expands when incorporating elements with larger atomic radii and shrinks when doped with elements possessing ionic radii smaller than that of calcium. In this systematic study, metal-doped calcium oxalate crystals were characterized by means of high-resolution synchrotron X-ray powder diffraction, energy-dispersive X-ray spectroscopy, inductively coupled plasma atomic emission spectroscopy, and transmission electron microscopy. Better understanding of metal incorporation into mineral hosts might lead to ways of developing new and more efficient sorbent materials for detoxification processes.

**Abbreviations**

| | |
|---|---|
| COM | calcium oxalate monohydrate |
| EDS | energy-dispersive X-ray spectroscopy |
| HRXRPD | high-resolution X-ray powder diffraction |
| SEM | scanning electron microscopy |
| TEM | transmission electron microscopy |
| HRTEM | high-resolution TEM |
| HAADF-STEM | high-angle annular dark-field scanning TEM |
| ICP-OES | inductively coupled plasma atomic emission spectroscopy |
| i.r. | ionic radius |

# 1. Introduction

Many metal contaminants are known to be toxic to the environment and to health. To minimize environmental pollution, attempts have been made to decontaminate soils, air, and water. One route is via the development of abundant, low-cost and effective materials that can be used to retain industrial effluent. Biological systems have turned out to be valuable entities for studying the impact of metals on the environment and for gaining a better understanding of the mechanisms involved in detoxification processes. As both biogenic and non-biogenic crystals have been shown to serve as hosts for organic molecules[1, 2] and for metal ions, their interaction has been extensively studied. In particular, calcium-bearing minerals that plant organisms produce,[3-10] as well as calcium carbonate and hydroxyapatite,[11-14] have been shown to absorb or exchange foreign cations effectively.[14-16]

A variety of cations have been tested for their sorption behavior especially on calcium carbonate and hydroxyapatite. Here we refer to some selected examples. In the case of calcium carbonate and hydroxyapatite, the adsorption affinity of $Cd^{2+}$ for these materials was found to be higher than that of $Zn^{2+}$.[14] The latter showed, in addition, non-linear adsorption behavior on calcite, leading the authors to suggest that $Zn^{2+}$ cations form a hydrated surface complex prior to dehydration. Based on this notion it was postulated that the incorporation of $Zn^{2+}$ into calcium carbonate is facilitated via recrystallization.[14] $Cd^{2+}$ cations, whose electronic configurations are comparable to those of $Ca^{2+}$, when exposed to geological calcite were shown to diffuse first into the surface layers and then into the bulk material.[12] Although different mechanisms were proposed for the interaction of Pb with calcite, Sturchio *et al.* observed that $Pb^{2+}$ occupies $Ca^{2+}$ sites in calcium carbonate.[17]

In contrast to the well-studied sorption phenomenon on calcium carbonate and hydroxyapatite, little is known about the interaction of metals with calcium oxalate. The structure of different calcium oxalate phases has been investigated in some theoretical and experimental studies[18, 19], and several studies have addressed the interaction of various organic molecules with calcium oxalate.[20-22] Research that led to the identification of organic modifiers as growth inhibitors of calcium oxalate[23, 24] attracted the attention of scientists because of its potential relevance for pharmaceutical development and crystal engineering. Nevertheless, relatively few studies have demonstrated the incorporation of metals into synthetic calcium oxalates, as shown, for example, in the case of $Sr^{2+}$.[25, 26] As calcium oxalate is rather insoluble above pH 3.5, it might be an advantage that long-term stability of the metal-doped phase occurs at a neutral pH.[25] Mechanisms suggested so far are based either on adsorption/desorption processes or on the replacement of $Ca^{2+}$ cations by other cations.[26] Clearly, however, a better understanding of the crystal structure of metal-doped calcium oxalate is needed.

In the framework of this study, we aimed to characterize the crystal structure of metal-doped calcium oxalate monohydrate (COM) and to gain a fundamental understanding of the mechanisms of incorporation of different metals. Achievement of a deeper understanding can be expected to open the way to the development of novel and more efficient sorbent materials that can be applied to detoxification processes.

## 2. Experimental Section
### 2.1 Calcium oxalate synthesis

Calcium oxalate was synthesized by simultaneously dropping 50 ml of 0.04 M $CaCl_2*2H_2O$ (Merck, Darmstadt, Germany) and 50 ml of 0.04 M $Na_2C_2O_4$ (Fluka, Seelze, Germany) solution into 200 ml of deionized (DI) water at 70 °C and stirring for one hour.[27] For the preparation of metal-doped COM, the metal salts [$CuSO_4*5H_2O$ (Merck, Darmstadt, Germany), $CdCl_2*2.5H_2O$ (Sigma-Aldrich, St. Louis, MO, USA), $PbCl_2$ (Spectrum, New Brunswick NJ, USA), $SrCl_2*6H_2O$ (BDH, Dawsonville GA, USA), and $ZnCl_2$ (Merck, Darmstadt, Germany)] were dissolved in preheated DI water to a final concentration of 100 μg/ml and a molarity of 0.40 mM for $CuSO_4*5H_2O$, 0.44 mM for $CdCl_2*2.5H_2O$, 0.36 mM for $PbCl_2$, 0.38 mM for $SrCl_2*6H_2O$, and 0.73 mM $ZnCl_2$. Precipitated crystals were filtered (Whatman #1005, Little Chalfont, UK), washed with DI water, and allowed to dry in air. All reagents were prepared in triplicates.

### 2.2 High-resolution X-ray powder diffraction

The HRXRPD pattern was obtained from dried calcium oxalate crystals measured in a borosilicate 1 mm capillary glass. Experiments were performed on the ID22 beamline of the European Synchrotron Radiation Facility (ESRF), Grenoble, France at a wavelength of 0.496006(9) Å. The Rietveld refinement method with GSAS-II software [28] was applied for structural analyses.

### 2.3 Scanning electron microscopy

Crystal shapes were determined using a SEM (Quanta 200, FEI, Hillsboro, OR, USA) at high voltage (15−20 kV). All samples were carbon coated prior to analysis (Quorum Q150T ES, East Grinstead, UK).

### 2.4 Transmission electron microscopy

TEM analyses were performed on a Titan FEG-S/TEM (FEI, Eindhoven, Netherlands) coupled with an EDS system for chemical analysis (Oxford Instruments, Oxfordshire, UK). All samples were ground to nanometer-sized powder and sprayed in ethanol on copper TEM grids. Micrographs were obtained in the HAADF-STEM mode.

### 2.5 Quantification of metals in calcium oxalate

Metal content and distribution were analyzed by energy-dispersive X-ray spectroscopy (EDS) mounted on a SEM, and by inductively coupled plasma atomic emission spectroscopy (ICP-OES, iCAP 6300 Duo, Thermo Scientific, Waltham, MA, USA). Using ICP-OES, calcium oxalate crystals (10 mg) were dissolved in 6 M HCl and diluted in deionized water up to a volume of 10 ml prior to measurement. The assumed error for ICP-OES measurements is 10 %.

## 3. Results and discussion

### 3.1 Morphology of synthetic metal-doped calcium oxalate monohydrate

Synthetic calcium oxalate monohydrate (COM) was prepared with the dual objective of investigating its ability to incorporate various metals and studying the effects of those metals on the COM crystal structure. COM crystals were precipitated in bulk assays in the absence and presence of $Pb^{2+}$, $Sr^{2+}$, $Cd^{2+}$, $Zn^{2+}$ and $Cu^{2+}$ salts, as described in the experimental section.

The crystal habit was characterized by scanning electron microscopy (SEM) and revealed that pure synthetic COM appears predominantly as twinned crystals (Figure 1) or as polycrystalline agglomerates (Figure S1, ESI). The presence of metal ions during crystal growth causes changes in the crystal morphology of COM compared to the pure COM reference sample. Whereas the appearance of the polycrystalline agglomerates is largely unaffected, the surfaces of twinned crystals appear distorted when doped with $Pb^{2+}$ (Figure 1B), $Sr^{2+}$ (Figure 1C), $Cd^{2+}$ (Figure 1D) or $Cu^{2+}$ (Figure 1F). In contrast to pure COM, moreover, $Pb^{2+}$ doping leads to both the formation of larger crystals and the growth of a more complex twinned morphology (Figure 1B), whereas the presence of $Sr^{2+}$ or $Cd^{2+}$ induces deformation of the COM habit, making the appearance of the twinned structure less obvious. $Zn^{2+}$-doped crystals have a roundish shape (Figure 1E). However, the crystals whose habits are most similar to the reference samples are those grown in a solution containing $Cu^{2+}$ (Figure 1F).

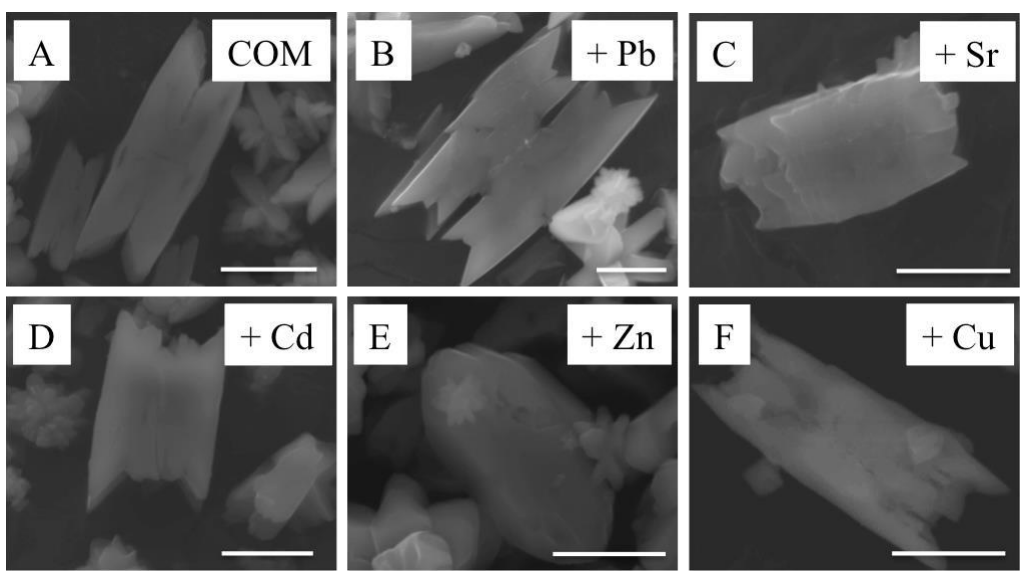

**Figure 1**. SEM micrographs of COM crystals precipitated in the absence (A) or in the presence of metal-containing solutions (B−F). Control crystals appear as a twin (A) or in a polycrystalline shape (Figure S1, ESI). Compared to the control sample (A), the crystal habits of metal-doped COM are distorted (B−F). (See Figure S1, ESI for overview images.) Scale bars, 5 μm.

### 3.2 Chemical characterization of metal-doped calcium oxalate monohydrate

To determine the quantity of metals incorporated into the COM phase we performed chemical analysis by energy-dispersive X-ray spectroscopy (EDS) using a SEM equipped with an EDS detector to study the crystal surface (to a depth of about 100 nm). To examine the bulk chemical composition we used inductively coupled plasma atomic emission spectroscopy (ICP-OES). These two techniques were used in a synergetic way to investigate cation-doping concentrations in the proximity of the crystal surface as well as within the crystal bulk (Figure 2). The EDS measurements indicated that $Cd^{2+}$, $Pb^{2+}$, $Sr^{2+}$ and $Zn^{2+}$ are included well within the host lattice, whereas $Cu^{2+}$ seems hardly to be incorporated at all. Relative to the EDS measurements, ICP-OES measurements yielded significantly lower atomic concentration values for Pb and Cd, but comparable values for Sr, Zn and Cu (Figure 2). The differences in metal concentrations obtained by the two methods, especially for Cd and Pb, might be explained by their sensitivities to surface (EDS) and bulk (ICP-OES) chemistry, which in our analysis indicated that Cd and Pb are enriched near the surface of the crystal. As Cu was detected only at very low concentration, we omitted the Cu-COM sample from the additional chemical analyses described in the following section.

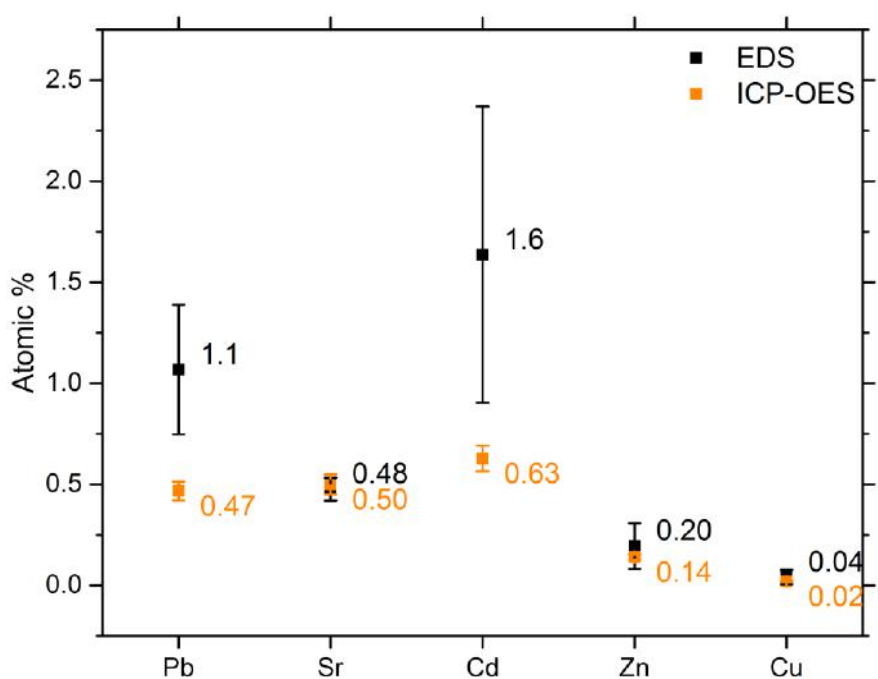

**Figure 2**. Atomic concentrations of metals detected in COM as determined by EDS point measurements (black squares) and ICP-OES (orange squares). EDS analysis revealed the following order of concentration: Cd (1.6 %) > Pb (1.1 %) > Sr (0.48 %) > Zn (0.20 %) > Cu (0.04 %); ICP-OES: Cd (0.63 %) > Sr (0.5 %) > Pb (0.47 %) > Zn (0.14 %) > Cu (0.02 %)

To better understand the distribution of metal cations within COM crystals, we performed line profile EDS chemical analysis of individual COM crystals of $Ca^{2+}$ and of crystals doped with three metal cations $Pb^{2+}$, $Sr^{2+}$, and $Cd^{2+}$ (Figure 3; black line = Ca, orange line = doping metal). The chemical composition indicates a non-homogeneous distribution of metal cations in the different samples. This effect is especially noticeable in the $Pb^{2+}$-doped phase, as can be seen from the line profile showing the highest concentration in the center of the crystal on its surface (Figure 3B, orange line). In contrast, for $Cd^{2+}$- and $Sr^{2+}$- doped phases the cations distribution is more uniform along the scanned area, as shown in Figure 3C, D.

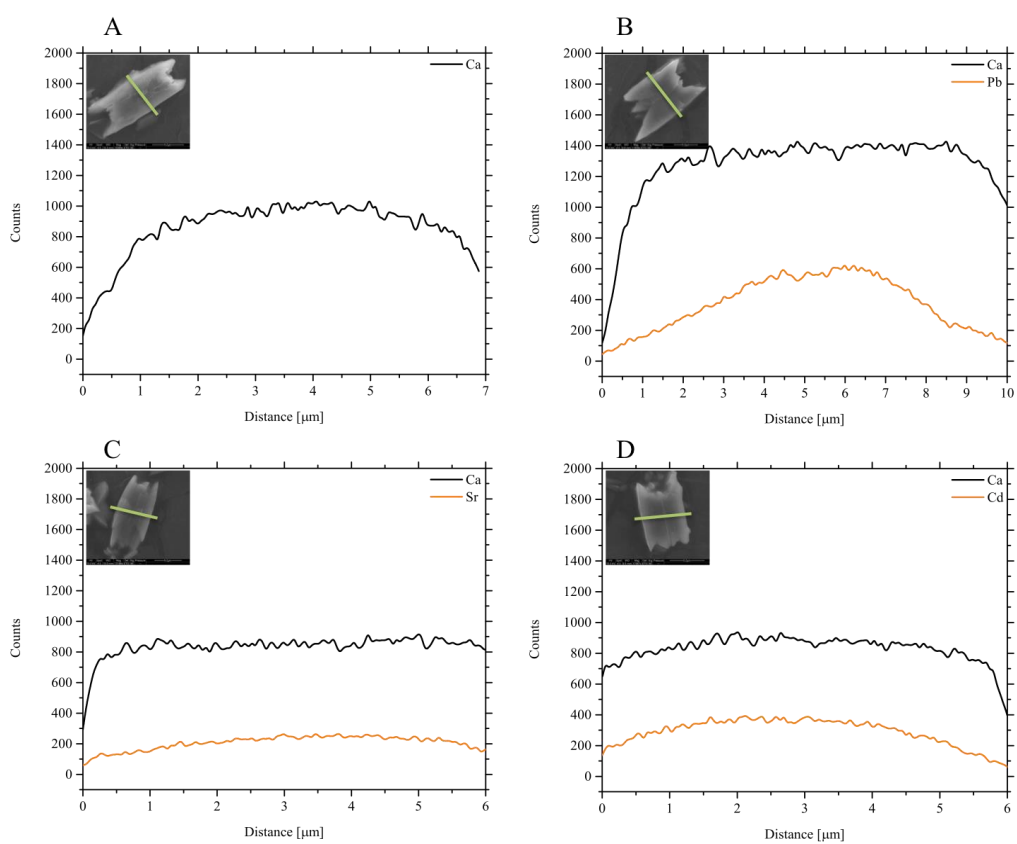

**Figure 3**. EDS analysis performed as line scan experiments on individual COM crystals (A) and COM crystals doped with $Pb^{2+}$ (B), $Sr^{2+}$ (C) and $Cd^{2+}$ (D) cations. Insets: SEM micrographs of the corresponding crystals. Green lines highlight line scan areas.

## 3.3 Visualization of metal cations within calcium oxalate monohydrate

Transmission electron microscopy (TEM) was further utilized to visualize the distribution of the selected cations $Cd^{2+}$, $Pb^{2+}$ and $Sr^{2+}$ at the nanoscale (Figure 4). In the high-angle annular dark-field scanning TEM mode (HAADF-STEM), the heavier elements appear brighter than the $Ca^{2+}$ cations because of the higher number of their electrons (z-contrast). The samples expose nanometric-sized brighter regions exclusively for $Pb^{2+}$- and $Cd^{2+}$-doped COM, indicating a local accumulation of these cations (Figure 4B, D). $Sr^{2+}$-doped COM also displays areas of brighter contrast, but the distribution is rather network-like, indicating a more homogeneous distribution of the cations (Figure 4C). Nevertheless, EDS confirmed the presence of metals in all areas presented in Figure 4 (Figure S2, ESI). Furthermore, the reference sample without doped metal cations did not show enhanced brightness in non-homogeneous regions.

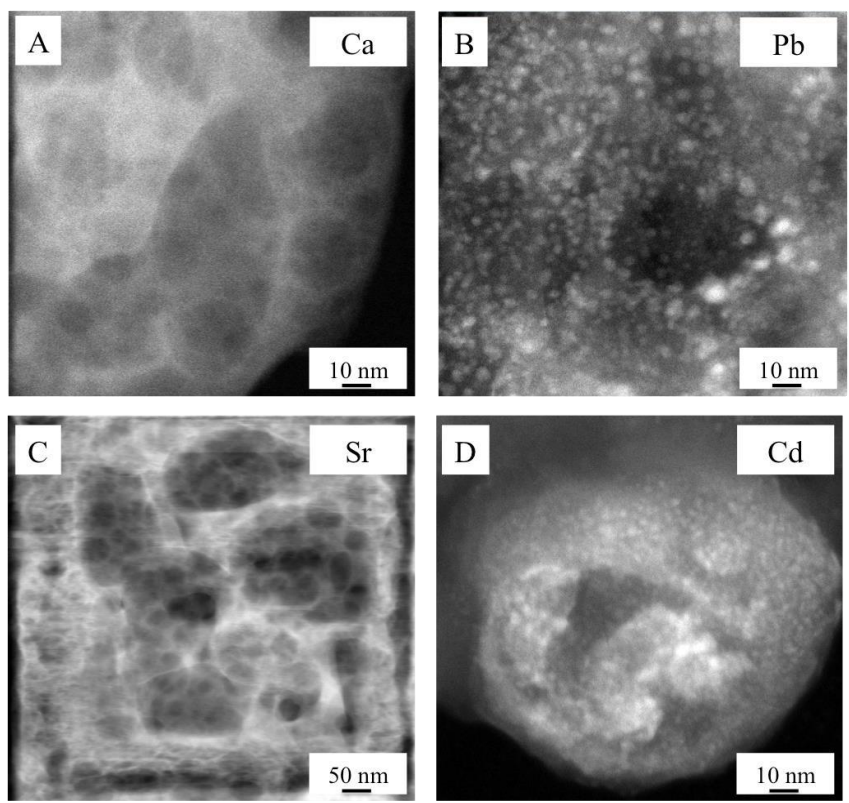

**Figure 4**. Transmission electron micrographs of pulverized COM crystals (COM reference (A), $Pb^{2+}$-doped COM (B), $Sr^{2+}$-doped COM (C), and $Cd^{2+}$-doped COM (D). $Cd^{2+}$ and $Pb^{2+}$ appear in COM as inclusions, as seen by the presence of spot-like

white areas in the micrographs. No such spots appear in the control (A). $Sr^{2+}$-doped COM shows a rather network-like distribution of brighter areas indicating its fairly homogeneous distribution within the material (C). EDS analysis confirmed the presence of individual elements for the areas presented here (Figure S2, ESI).

### 3.4 X-ray diffractometric analysis of metal-doped COM

We carried out high-resolution X-ray powder diffraction (HRXRPD) measurements using synchrotron light to correlate the differences in ionic radius of the selected metal cations with the previous observations on their chemical distributions in COM crystals. In addition, we wanted to understand how the foreign metals substitute for $Ca^{2+}$ within the lattice and to what extent the lattice parameters of the COM are distorted by such a substitution.

All of the collected diffraction patterns analyzed in this study are shown in Figure 5. Observed diffraction peaks can be attributed to the COM phase with the exception of a low-intensity peak at $2\theta = 7.023°$ (Figure 5A, arrow). By analyzing each diffraction peak more detailed, and comparing the diffraction peak positions of the doped and non-doped COM, we observed systematic shifts exclusively for $Pb^{2+}$, $Sr^{2+}$, $Cd^{2+}$ and $Zn^{2+}$-doped COM. These findings are presented here for the (100) and (040) reflections (Figure 5B) and confirm that $Ca^{2+}$ is indeed replaced by cations with different ionic radii. The only exception is displayed by the diffractogram of the Cu-COM phase that does not show any peak shift, demonstrating that $Cu^{2+}$ is hardly incorporated, if at all, into the structure of COM. This finding was confirmed by the chemical analysis showing a very low atomic concentration of Cu atoms in COM (Figure 2).

Furthermore, the direction of the peak shift indicates clear dependence of the ionic radius of the incorporated metal on the lattice parameter of the crystal host after its incorporation into the crystal lattice. According the Shannon database[29] the ionic radius for the used cations in six-fold coordination are: for $Ca^{2+}$ i.r. = 1 Å, for $Pb^{2+}$ i.r. = 1.19 Å, for $Sr^{2+}$ i.r. = 1.18 Å, for $Cd^{2+}$ i.r. = 0.95 Å, for $Zn^{2+}$ i.r. = 0.73 Å and for $Cu^{2+}$ i.r. = 0.73 Å. For the cations with larger ionic radii than that of $Ca^{2+}$, such as $Pb^{2+}$ and $Sr^{2+}$, the peaks are shifted to lower Bragg diffraction angles (larger $d$-spacings), whereas cations with smaller ionic radii, such as $Cd^{2+}$ and $Zn^{2+}$, induce the opposite effect. Reflection peaks in the $Cd^{2+}$- $Sr^{2+}$- and $Pb^{2+}$-doped COM are strongly asymmetric, highlighting the non-homogeneity of incorporation into the crystal in these samples. The material in fact contains zones in which the quantity of cation dopant is higher than the average, which is in good agreement with the electron microscopy data shown above (Figure 4).

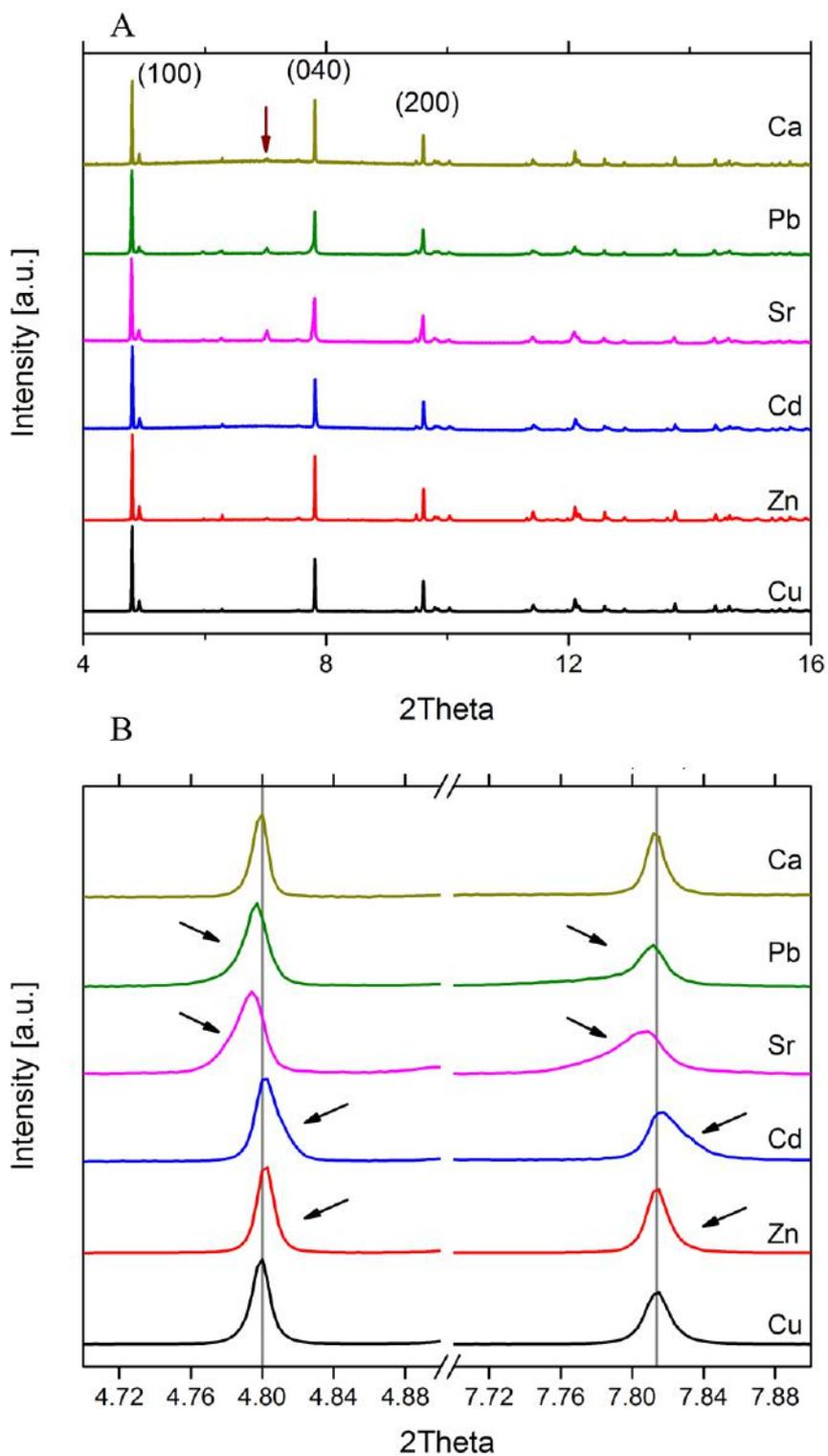

**Figure 5**. HRXRPD pattern of COM and metal-doped COM (A, full diffractogram; note that the single peak at 7.023 degrees could not be attributed to COM (dark-red arrow). Zooming in of the (100) and (040) reflections reveals a shift in the peak positions of doped and non-doped COM (B).

To quantify the cell stress in doped samples and to determine if the insertion of a small amount of dopant cations in COM can distort the atomic structure, a full structural Rietveld refinement was performed on all diffractograms of individual samples based on the structure determined by Daudon.[30] As an example, the plot of experimental and calculated curves of the reference COM sample is shown in Figure 6. Refined lattice constants for all samples and accordingly calculated lattice distortions along the entire principal axes ($\Delta a/a$, $\Delta b/b$, $\Delta c/c$) are reported in Figure 7 and Table S1, ESI.

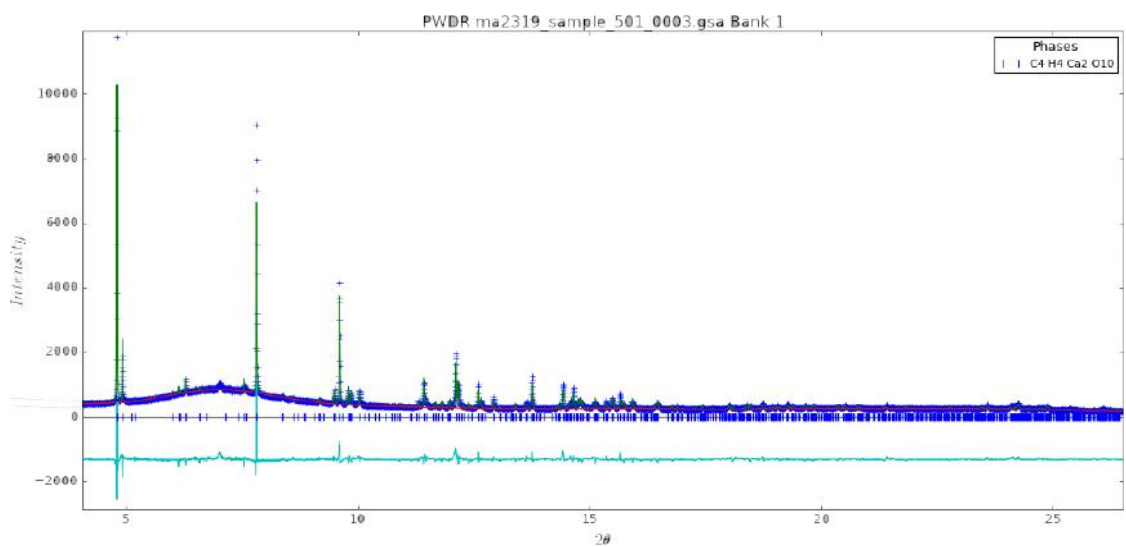

**Figure 6**. An example of the Rietveld refinement method applied to the COM control sample.

As shown in Figure 7, the results indicate that cations larger than $Ca^{2+}$ affect mainly the *a,b* lattice parameters ($Pb^{2+}$, $Sr^{2+}$), whereas smaller cations ($Cd^{2+}$, $Zn^{2+}$) influence preferentially the *c* lattice parameter of COM and not the *a* and *b* planes. Although both $Pb^{2+}$ and $Sr^{2+}$ are larger than $Ca^{2+}$, both cations interact in a distinct way with the pure phase. $Pb^{2+}$ seems to interact exclusively with planes perpendicular to *a* and *b*, as the lattice distortion along the *c* parameter is very small, with $\Delta c/c$ = -9.61E-05. $Sr^{2+}$, on the other hand, induces distortions along all crystallographic direction with $\Delta a/a$ =

1.58E-03, $\Delta b/b$ = 1.51E-03, and $\Delta c/c$ = 9.85E-04. Of the relatively smaller cations, $Cd^{2+}$ induced the most prominent negative distortions, with $\Delta a/a$ = -5.22E-04, $\Delta b/b$ = -6.18E-04 and $\Delta c/c$ = -7.88E-04.

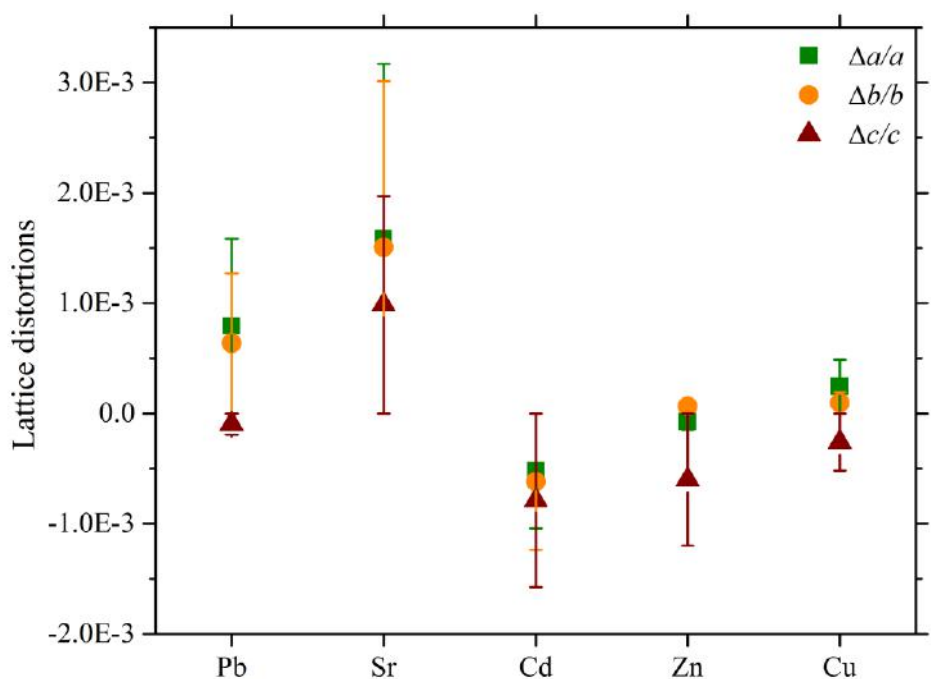

**Figure 7**. Lattice distortions are determined along the principal axes. The largest lattice distortions are observed along the *a,b* directions for cations with a radius larger than that of $Ca^{2+}$ ($Pb^{2+}$ and $Sr^{2+}$) whereas cations smaller than $Ca^{2+}$ seem to induce lattice distortions preferentially along the *c*-direction ($Cd^{2+}$, $Zn^{2+}$ and $Cu^{2+}$).

The diffraction measurements further showed that the dimensions of the unit cells are distorted when the COM is precipitated in the presence of foreign divalent metal cations, with cations having larger ionic radii than $Ca^{2+}$ leading to expansion and cations with smaller ionic radii than $Ca^{2+}$ inducing shrinkage of the COM unit cell. This finding led us to expect that the volume of the corresponding unit cells is also changed accordingly. To verify this hypothesis we compared the volumes of individual unit cells of COM and metal-doped COM as retrieved by the Rietveld refinement method (Figure 8). The volume of the reference COM unit cell was found to be 870.304 $Å^3$. For COM doped with larger cations than $Ca^{2+}$ we observed an increase in the unit cell volume to 871.329 $Å^3$ (0.12 %) for $Pb^{2+}$ and 873.706 $Å^3$ (0.39

%) for $Sr^{2+}$ whereas cations smaller than $Ca^{2+}$ led to a reduction in the unit cell volume: 868.475 Å³ (−0.21 %) for $Cd^{2+}$, 869.525 Å³ (−0.09 %) for $Zn^{2+}$ and 870.289 Å³ (−0.002 %) for $Cu^{2+}$. These findings are in good agreement with our hypothesis.

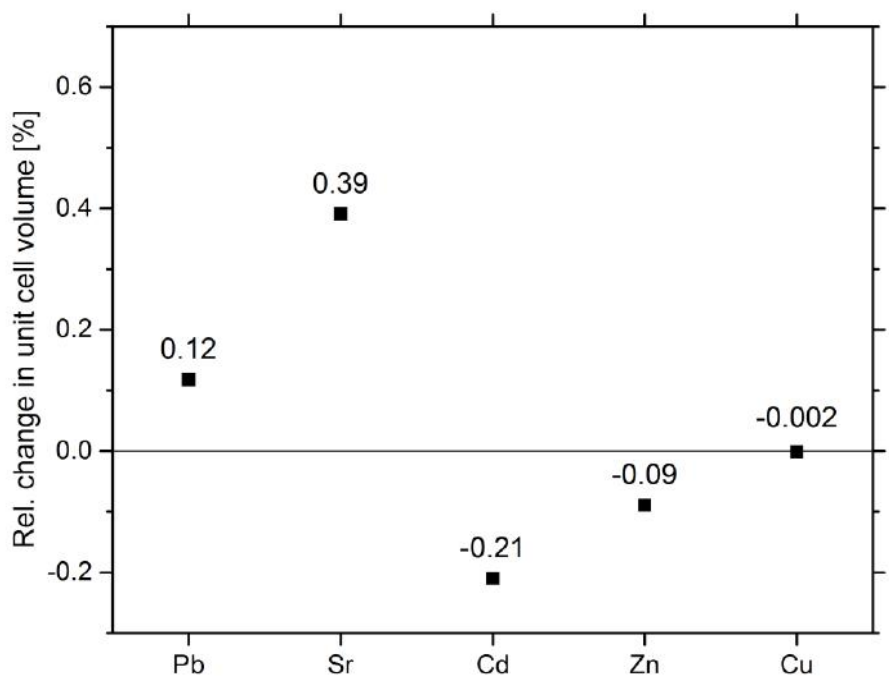

**Figure 8.** Relative changes in unit cell volume calculated for metal-doped COM based on values determined by the Rietveld refinement method.

To correlate the change in the unit cell dimensions with the amount of incorporated foreign metal cations, we calculated the weighted change in lattice dimensions based on the average ionic radius and the quantity of incorporated cations.

i.r. (average) = (at %(Ca) × i.r.(Ca)) + (at %(dopant) × i.r.(dopant))

(i.r., ionic radius)

The results are shown for each crystallographic direction (Figure 9A, B and C). The plots highlight a non-linear dependence of the weighted average radius on the induced lattice distortions. An anomaly is apparent in the lattice distortion behavior for $Pb^{2+}$- and $Sr^{2+}$-doped COM in that these cations have the same average ionic radii but different lattice parameters. This phenomenon might be explained by some characteristics of $Sr^{2+}$. First, $Sr^{2+}$ has a strong chemical affinity for $Ca^{2+}$ and $Ca^{2+}$ can be easily replaced by $Sr^{2+}$ in many phases; secondly, $Pb^{2+}$ is reported to have a lone pair of electrons and thus, unlike $Sr^{2+}$ and $Ca^{2+}$, cannot be modeled as spherical. This anisotropic character of the $Pb^{2+}$ cation might explain why it interacts with the *a* and *b* axis in preference to the *c* axis. Moreover, chemical analysis shows that $Pb^{2+}$ cations are located mainly at the crystal surface, whereas $Sr^{2+}$ cations are equally distributed near the surface and the bulk of the crystal. All of these elements probably contribute to the generation of this anomaly in the lattice parameters.

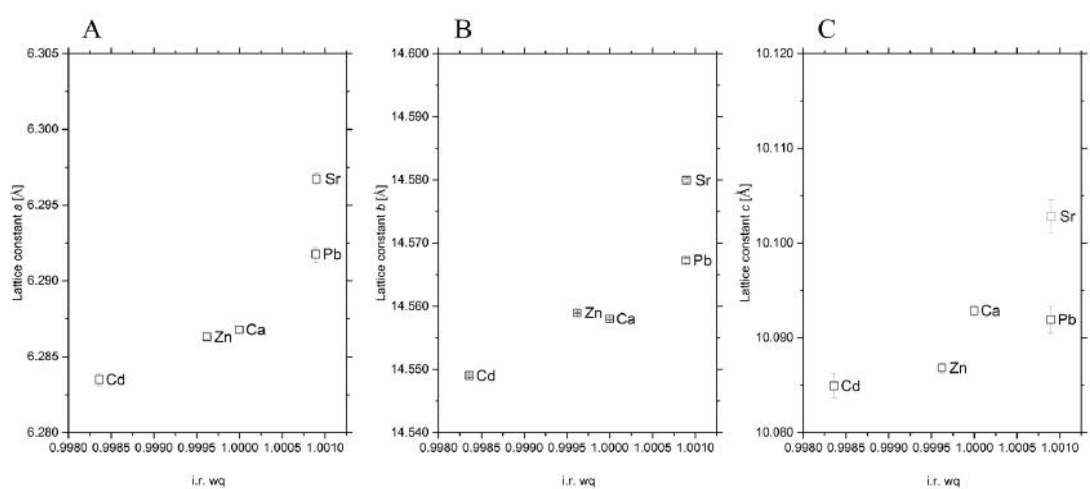

**Figure 9.** Lattice constants of metal-doped COM, shown relative to the weighted quantity of cations in the COM structure. Data are shown for the main crystallographic directions *a* (A), *b* (B), *c* (C). (i.r. wq = ionic radius weighted to the quantity of incorporation)

Aiming a better understanding of the unit cell structure of doped COM, we calculated the distances for Ca-O in each sample and the results show that the values are scattered around an average of 2.46(3) Å for Ca1-O and 2.40(3) Å for Ca2-O (Table S2, ESI). This might be explained due to sample inhomogeneity, where we observe a peak broadening that puts a limit towards more precise structure refinements and calculations.

**4. Conclusions**

In this study, we doped COM with different divalent metal cations ($Pb^{2+}$, $Sr^{2+}$, $Cd^{2+}$, $Zn^{2+}$, and $Cu^{2+}$) to determine the suitability of each of these cations for entry into the structure of COM and to examine how the COM lattice parameters are affected by their incorporation. Considering the Goldschmidt´s rules for element substitution in crystals that encompasses ionic radius, ionic charge, ionic potential and electronegativity, we could expect that $Cd^{2+}$ is the most appropriate candidate for substitution according to its relative size and $Sr^{2+}$ according to its relative electronegativity. Moreover, $Zn^{2+}$ and $Cu^{2+}$ would be less suitable candidates for the substitution of $Ca^{2+}$ as their ionic radii differ the most from all the other investigated ions and both cations exhibit a rather high electronegativity compared to $Ca^{2+}$.

Based on chemical analyses, x-ray powder diffraction and high-resolution TEM imaging, we clearly showed that $Ca^{2+}$ can be replaced in the lattice by metals, when the crystal host was grown in the presence of $Pb^{2+}$, $Sr^{2+}$, $Cd^{2+}$, or $Zn^{2+}$. We found that $Cu^{2+}$ is scarcely incorporated into the COM and the $Cu^{2+}$-doped COM crystals do not exhibit significant lattice parameter change with respect to that of the reference sample. In addition to $Cu^{2+}$, the concentration of $Zn^{2+}$ was also found to be lower compared to the other investigated cations, but slightly higher than that of $Cu^{2+}$. This is surprising as both have a very similar ionic radius. One possible explanation to this can be derived via the Goldschmidt´s substitution rule that also relates to the relative electronegativity: the electronegativity of $Zn^{2+}$ is lower than that of $Cu^{2+}$ which means that the Zn-O bond is more ionic like than that of Cu-O which is closer to that of Ca-O and may indeed explain the difference in incorporation levels.

Moreover, we observed a general trend towards correlation of the lattice distortions with the ionic radii of the tested metals. When cations larger than $Ca^{2+}$ (1 Å) were incorporated the unit cells were found to expand, whereas with cations smaller than $Ca^{2+}$ the unit cells tended to shrink. In addition, we found that when the ionic radius of the incorporated cation is larger than 1 Å the lattice distortions occur preferentially along the *a,b* direction, whereas when the ionic radius is smaller than 1 Å the distortion is mainly along the *c*-axis.

Despite this general trend, each individual cation was found to interact with COM in its own specific way. Of the cations used in this study $Sr^{2+}$ is the most similar to $Ca^{2+}$, and therefore becomes incorporated by its substitution into the lattice in the highest amounts. $Pb^{2+}$, whose ionic radius is the largest, was detected in relatively high concentrations on the sample surface. Although the ionic radii of $Pb^{2+}$ and $Sr^{2+}$ cations are similar, the lattice distortions induced by $Pb^{2+}$ are lower than those induced by $Sr^{2+}$. Incorporation of $Pb^{2+}$ into the crystal lattice of COM seemed to be somewhat problematic and the interaction of this cation with the crystal surface might be preferred. As the ionic radii of $Cd^{2+}$ (0.95 Å) and $Zn^{2+}$ (0.73 Å) are smaller $Ca^{2+}$ (1 Å) the lattice distortions with these cation were negative as expected. Nevertheless, as the diffraction peaks of the $Zn^{2+}$-doped COM were found to be the most symmetrical of all other doped sample tested in this work, this might indicate that its incorporation is more homogeneous than that of $Pb^{2+}$, $Sr^{2+}$ or $Cd^{2+}$-doped samples.

It is noteworthy to mention that in two recent studies authors report on the formation of amorphous calcium oxalate *in vitro*.[31, 32] Here, although there is no experimental evidence, we cannot exclude that a minor amorphous phase appears locally due to the doping and local distortions. A further characterization by means of X-ray absorption spectroscopy[26] and selected area electron diffraction experiments might help to better understand the chemical environment of doped ions as well as provide more information on the local structure. However, the substitution of ions in a crystal lattice is a complex process and various parameters need to be considered. Beyond the physico-chemical properties of the cations, the experimental setup may also influence the doping procedure such as reaction temperature and dopant concentration in growth solution.

Although open questions remain, we could show that growing crystals of COM interact with various divalent metal cations, so that the latter are potential candidates to be considered by substitution of $Ca^{2+}$ in various detoxification procedures.

**Acknowledgement**

The research leading to these results received funding from the European Research Council under the European Union's Seventh Framework Program (FP/2007-2013)/ERC Grant Agreement (no. 336077). We also thank the Minerva foundation for financial support. The authors gratefully appreciate the help of Yaron Kauffman from the MIKA center for TEM measurements, and of Malik Kochva, Agricultural Engineering Department at the Technion−Israel Institute of Technology for conducting the ICP-OES measurements.

# References

1. S. Borukhin, L. Bloch, T. Radlauer, A. H. Hill, A. N. Fitch and B. Pokroy, *Adv Funct Mater*, 2012, **22**, 4216-4224.
2. E. Weber and B. Pokroy, *Crystengcomm*, 2015, **17**, 5873-5883.
3. V. R. Franceschi and P. A. Nakata, *Annual Review of Plant Biology*, 2005, **56**, 41-71.
4. E. Zindler-Frank, *Botanica Acta*, 1991, **104**, 229–232.
5. H. H. He, T. M. Bleby, E. J. Veneklaas, H. Lambers and J. Kuo, *Plos One*, 2012, **7**.
6. E. Van Balen, S. C. Van de Geijn and G. M. Desmet, *Zeitschrift für Pflanzenphysiologie*, 1980, **97**, 123 -133.
7. A. M. A. Mazen and O. M. O. El Maghraby, *Biol Plantarum*, 1997, **40**, 411-417.
8. F. Faheed, A. Mazen and S. Abd Elmohsen, *Turk J Bot*, 2013, **37**, 139-152.
9. M. P. Isaure, G. Sarret, E. Harada, Y. E. Choi, M. A. Marcus, S. C. Fakra, N. Geoffroy, S. Pairis, J. Susini, S. Clemens and A. Manceau, *Geochim Cosmochim Ac*, 2010, **74**, 5817-5834.
10. G. Sarret, E. Harada, Y. E. Choi, M. P. Isaure, N. Geoffroy, S. Fakra, M. A. Marcus, M. Birschwilks, S. Clemens and A. Manceau, *Plant Physiol*, 2006, **141**, 1021-1034.
11. A. Garcia-Sanchez and E. Alvarez-Ayuso, *Miner Eng*, 2002, **15**, 539-547.
12. S. L. Stipp, M. F. Hochella, G. A. Parks and J. O. Leckie, *Geochim Cosmochim Ac*, 1992, **56**, 1941-1954.
13. J. M. Zachara, C. E. Cowan and C. T. Resch, *Geochim Cosmochim Ac*, 1991, **55**, 1549-1562.
14. J. A. Gomez del Rio, P. J. Morando and D. S. Cicerone, *J Environ Manage*, 2004, **71**, 169-177.
15. A. Bigi, G. Falini, E. Foresti, M. Gazzano, A. Ripamonti and N. Roveri, *J Inorg Biochem*, 1993, **49**, 69-78.
16. A. Bigi, G. Falini, M. Gazzano, N. Roveri and E. Tedesco, *Mater Sci Forum*, 1998, **278-2**, 814-819.
17. N. C. Sturchio, R. P. Chiarello, L. W. Cheng, P. F. Lyman, M. J. Bedzyk, Y. L. Qian, H. D. You, D. Yee, P. Geissbuhler, L. B. Sorensen, Y. Liang and D. R. Baer, *Geochim Cosmochim Ac*, 1997, **61**, 251-263.
18. O. Hochrein, A. Thomas and R. Kniep, *Z Anorg Allg Chem*, 2008, **634**, 1826-1829.
19. A. Millan, *Cryst Growth Des*, 2001, **1**, 245-254.
20. K. R. Cho, E. A. Salter, J. J. De Yoreo, A. Wierzbicki, S. Elhadj, Y. Huang and S. R. Qiu, *Cryst Growth Des*, 2012, **12**, 5939-5947.
21. B. Grohe, K. A. Rogers, H. A. Goldberg and G. K. Hunter, *J Cryst Growth*, 2006, **295**, 148-157.
22. B. Grohe, S. Hug, A. Langdon, J. Jalkanen, K. A. Rogers, H. A. Goldberg, M. Karttunen and G. K. Hunter, *Langmuir*, 2012, **28**, 12182-12190.
23. S. Farmanesh, J. Chung, D. Chandra, R. D. Sosa, P. Karande and J. D. Rimer, *J Cryst Growth*, 2013, **373**, 13-19.
24. S. Farmanesh, S. Ramamoorthy, J. H. Chung, J. R. Asplin, P. Karande and J. D. Rimer, *J Am Chem Soc*, 2014, **136**, 367-376.
25. S. Chatterjee, Temple University, 2009.

26. D. M. Singer, S. B. Johnson, J. G. Catalano, F. Farges and G. E. Brown, *Geochim Cosmochim Ac*, 2008, **72**, 5055-5069.
27. N. Bouropoulos, S. Weiner and L. Addadi, *Chem-Eur J*, 2001, **7**, 1881-1888.
28. B. H. Toby and R. B. Von Dreele, *Journal of Applied Crystallography*, 2013, **46**, 544-549.
29. R. D. Shannon, *Acta Crystallogr A*, 1976, **32**, 751-767.
30. M. Daudon, D. Bazin, G. Andre, P. Jungers, A. Cousson, P. Chevallier, E. Veron and G. Matzen, *Journal of Applied Crystallography*, 2009, **42**, 109-115.
31. J. Ihli, Y.-W. Wang, B. Cantaert, Y.-Y. Kim, D. C. Green, P. H. H. Bomans, N. A. J. M. Sommerdijk and F. C. Meldrum, *Chem Mater*, 2015, **27**, 3999-4007.
32. M. Hajir, R. Graf and W. Tremel, *Chem Commun*, 2014, **50**, 6534-6536.